\documentclass[11pt]{article}

\usepackage{amssymb}
\usepackage[dvips]{graphicx}
\usepackage{citesort}

\def\bra#1{\left\langle#1\right|}
\def\ket#1{\left|#1\right\rangle}

\begin{document}
\vspace*{1cm}
\begin{center}
{\LARGE\bf\sf Proving the AGT relation for $N_{\!f}=0,1,2$  antifundamentals} \\
\end{center}

\begin{center}

\vspace*{2cm}

    {\large\bf\sf
    Leszek Hadasz${}^\dag$\footnote{\emph{e-mail}: hadasz@th.if.uj.edu.pl}$\!\!\!\!,\ \,$
    Zbigniew Jask\'{o}lski${}^\ddag$\footnote{\emph{e-mail}: jask@ift.uni.wroc.pl}
    and
    Paulina Suchanek${}^\ddag$\footnote{\emph{e-mail}: paulina@ift.uni.wroc.pl}
    }
     \\
\vskip 3mm
    ${}^\dag$ M. Smoluchowski Institute of Physics,
    Jagiellonian University \\
    Reymonta 4,
    30-059~Krak\'ow, Poland, \\

\vskip 3mm
    ${}^\ddag$ Institute of Theoretical Physics,
    University of Wroc{\l}aw \\
    pl. M. Borna 9, 50-204~Wroc{\l}aw, Poland. \\
\end{center}

\vspace*{1cm}
\begin{abstract}
Using recursive relations  satisfied by Nekrasov partition functions and by irregular conformal blocks we prove
the AGT correspondence in the case of ${\cal N}=2$ superconformal SU(2) quiver gauge theories with $N_{\!f}=0,1,2$
antifundamental hypermultiplets.

\end{abstract}

\vspace*{\fill}

PACS: 11.25.Hf, 11.30.Pb

\section{Introduction}

Last year Alday, Gaiotto and Tachikawa conjectured \cite{Alday:2009aq}   that
partition functions of ${\cal N}~=~2$ superconformal SU(2) quivers are directly related to
correlation functions of the two dimen\-sio\-nal Liouville field theory.
This relation has been soon extended to similar relations between the general SU($N$) quiver theories and
$A_{N-1}$ Toda theories \cite{Wyllard:2009hg}   and to other objects like surface and loop operators
and their Liouville counterparts \cite{Drukker:2009tz,Alday:2009fs,Drukker:2009id,Gaiotto:2009fs,Wu:2009tq,Passerini:2010pr}.
Other extensions concern non-conformal limit of the AGT \cite{Gaiotto:2009ma,Marshakov:2009gn,Alba:2009fp,Taki:2009zd,Yanagida:2010kf}
correspondence and its 5-dimensional version \cite{Awata:2009ur,Yanagida:2010vz}.
Yet another generalization has been recently found in \cite{Santachiara:2010bt}.

An explanation of the AGT relation was given by Dijkgraaf and Vafa \cite{Dijkgraaf:2009pc}. The idea was
to relate  both sides of the correspondence to a certain class of matrix models.
These relations were further analyzed in a number of papers
\cite{Itoyama:2009sc,Eguchi:2009gf,Schiappa:2009cc,Mironov:2009ib,Sulkowski:2009ne,Shakirov:2009nx,Fujita:2009gf}.
Another M-theory explanation was presented in
\cite{Bonelli:2009zp,Alday:2009qq}.

An essential part of the AGT conjecture
is an exact correspondence between  instanton parts of  Nekrasov
partition functions in 4-dimensional ${\cal N} =2$ SCFT  \cite{Nekrasov:2002qd} and  conformal blocks of the 2-dimensional CFT \cite{Belavin:1984vu}.
This relation has passed many checks
\cite{Marshakov:2009gs,Mironov:2009dr,Mironov:2009qt,Mironov:2009by,Mironov:2009qn,Giribet:2009hm,Alba:2009ya}
and lead to many interesting results both on the conformal blocks
and on the ${\cal N}=2$ quivers
\cite{Nanopoulos:2009au,Marshakov:2009kj,Poghossian:2009mk,Mironov:2009uv,Nanopoulos:2009uw,Hadasz:2009db,Hadasz:2009sw,Mironov:2009dv,Kanno:2009ga,Fateev:2009aw,Petkova:2009pe,Mironov:2010zs,Popolitov:2010bz,Itoyama:2010ki,Mironov:2010ym,Nekrasov:2010ka,Drukker:2010jp}.

In spite of all these developments only in the case
of ${\cal N}=2$ SU(2) SYM with a single adjoint matter multiplet an analytic proof of the AGT relation is known.
The main idea of the  proof given by Fateev and Litvinov \cite{Fateev:2009aw} is to show that the corresponding
Nekrasov function and the 1-point conformal block on the torus satisfy exactly the same
recursion relations. These relations  were first conjectured by Poghossian \cite{Poghossian:2009mk}
  and then  proven on the CFT side in \cite{Hadasz:2009db} and on the ${\cal N}=2$ SU(2) SYM side in \cite{Fateev:2009aw}.
The aim of the present paper is to extend this proof to the case of
 $N_{\!f}=0,1,2$
antifundamental hypermultiplets.

In Section 2 we use the results of Marshakov, Mironov and Morozov \cite{Marshakov:2009gn} and Poghossian \cite{Poghossian:2009mk}
to derive the recursive relations for all irregular blocks by analyzing appropriate decoupling limits
of the Zamolodchikov elliptic recursive relation for the 4-point conformal block on the sphere
\cite{Zamolodchikov:ie,Zamolodchikov:2,Zamolodchikov:3}. This allows in particular to prove the relation
between two representations of the irregular block with two $\mu$ parameters conjectured in \cite{Gaiotto:2009ma}.
As a side topic we clarify the relations of polynomials
appearing in the recursive relations to the fusion polynomials and to the null states in the degenerate
Verma modules. These new, intriguing result deserves further investigations.

In Section 3 we follow the method of Fateev and Litvinov \cite{Fateev:2009aw}
 to analyze singularities of the Nekrasov functions and the factorization of the residues in the case of an arbitrary number of antifundamentals.
The complete  derivation of recursive formulae along this line requires the large
 $ {\sf p} = {a\over \hbar}$ asymptotic of the Nekrasov functions. This is simple in the cases $N_f=0,1$. We were
 also able
to calculate the asymptotic in the $N_f=2$ case.
In all three cases the  recursions obtained are identical  on both sides of the correspondence.
Calculating the asymptotics  in the cases $N_f=3,4$ turned out to be difficult and is still a challenging open problem.

\section{Recursive relations for irregular blocks}

The Gaiotto states \cite{Gaiotto:2009ma} can be defined by the conditions
\begin{eqnarray*}
L_1 \ket{\Delta, \Lambda^2} &=& -\Lambda^2  \ket{\Delta, \Lambda^2},
\qquad
L_n \ket{\Delta, \Lambda^2} = 0 \quad \mathrm{for}\quad n \geq 2.
\\
L_2 \ket{\Delta,\mu ,\Lambda} &=&- \Lambda^2  \ket{\Delta,\mu ,\Lambda},
\\
L_1 \ket{\Delta,\mu ,\Lambda} &=& -2\mu  \Lambda  \ket{\Delta,\mu ,\Lambda},
\qquad
L_n \ket{\Delta,\mu ,\Lambda} = 0 \quad \mathrm{for}\quad n \geq 3.
\end{eqnarray*}
An explicit form of these states was found  in \cite{Marshakov:2009gn}
\begin{eqnarray}\label{state1}
\ket{\Delta, \Lambda^2} &=& \sum_{n=0} \Lambda^{2n} \ket{\Delta,n} =
 \sum_n \Lambda^{2n}\, (-1)^n \sum_{|J|=n}  \left[B^{n}_{c, \Delta}\right]^{[1^{n}],J} \, L_{-J} \ket{\Delta},
\\
\label{state2}
\ket{\Delta,\mu ,\Lambda} &=& \sum_{n=0}  \Lambda^{n} \ket{\Delta,\mu ,n}
 \\ \nonumber
&=&
 \sum_n \Lambda^{n} \, \sum_{|J|=n} \,
 \sum_{p=0}^{n/2} \,(-1)^{n-p} (2\mu )^{n-2p} \left[B^{n}_{c, \Delta}\right]^{[1^{n-2p},2^p],J} \, L_{-J} \ket{\Delta}.
\end{eqnarray}
In the formulae above  $\displaystyle\left[B^{n}_{c, \Delta}\right]^{I,J}$ denotes the inverse of
the Gram matrix
$$
\left[B^{n}_{c, \Delta}\right]_{I,J}= \bra{\Delta} L_I L_{-J} \ket{\Delta}
$$
in the standard basis
$$
L_{-J} \ket{\Delta}= L_{-j_1}\dots L_{-j_k} \ket{\Delta},\;\;\;j_1 \leqslant \dots \leqslant j_k\,,\;\;|J|=\sum\limits_{i=1}^k j_i
$$
of the Verma module ${\cal V}_{c,\Delta}$ of the central charge $c$ and the highest weight $\Delta$.

The irregular blocks \cite{Gaiotto:2009ma} are defined as scalar products of the Gaiotto states:
\begin{eqnarray*}
\left\langle \Delta, \Lambda^2  | \Delta,\Lambda^2  \right\rangle
&=&
 \sum_{n} \Lambda^{4n}  \left\langle \Delta,  n \, |\,\Delta, n \right\rangle,
 \\ [4pt]
\left\langle \Delta,  \mu ,{\textstyle \frac{1}{2}}\Lambda\, |\,
\Delta, \Lambda^2 \right\rangle
&=&
\sum_n \Lambda^{3n} 2^{-n}
\left\langle \Delta, \mu, n \, |\,\Delta, n \right\rangle,
\\[4pt]
\left\langle \Delta, \mu _1,{\textstyle \frac{1}{2}}\Lambda\, |\, \Delta, \mu _2, {\textstyle \frac{1}{2}}\Lambda\right\rangle
&=&
\sum_n \Lambda^{2n}2^{-2n}
\left\langle \Delta, \mu_1, n \, |\,\Delta, \mu_2,n \right\rangle,
\end{eqnarray*}
or in terms of the 3-point conformal block\footnote{
The present notation for 3-point conformal block is related to that used in \cite{Hadasz:2009db,Hadasz:2006sb}
by
$
\bra{\xi'}V_\Delta(1)\ket{\xi''}=\rho(\xi',\nu_{\Delta},\xi'') .
$
The normalization condition takes the form
$\bra{\Delta'}V_\Delta(1)\ket{\Delta''}=1$.
}:
\begin{eqnarray}
\label{B block}
\bra{\Delta,\Lambda^2}V_{\Delta_2}(1)\ket{\Delta_1}
&=&
\sum\limits_{n} \Lambda^{2n}\bra{\Delta,n}V_{\Delta_2}(1)\ket{\Delta_1},
\\
\nonumber
\bra{\Delta,  \mu_3 ,{\textstyle \frac{1}{2}}\Lambda}V_{\Delta_2}(1)\ket{\Delta_1}
&=&
\sum\limits_{n} \Lambda^{2n}2^{-2n}\bra{\Delta,\mu_3, n}V_{\Delta_2}(1)\ket{\Delta_1},
\end{eqnarray}
where the conformal weights are related to $\mu_i$ parameters by
\begin{equation}
\label{parameters1}
\Delta_i = \frac14\left(Q^2 - \lambda_i^2\right),\;\;\;\;\;\;
2\mu_1 = \lambda_2+\lambda_1\,
,\;\;\;\;\;\;
2\mu_2 = \lambda_2-\lambda_1\,.
\end{equation}
It was shown in \cite{Marshakov:2009gn} that all irregular blocks above can be obtained by
appropriate decoupling limits of the  4-point conformal block on the sphere :
$$
\mathcal{B}_{\Delta}\! \left[^{ \Delta_3 \; \Delta_2}_{\Delta_4 \;\Delta_1} \right]\!(x)
= \sum_{n=0} x^n
\sum\limits_{|J|=|K|=n}
\bra{\Delta_4}V_{\Delta_3}(1)L_{-J}\ket{\Delta}
\left[ B^{\,{n}}_{c,\Delta}\right]^{J,K}
\bra{\Delta}L_K V_{\Delta_2}(1)\ket{\Delta_1}.
$$
If
\begin{equation}
\label{parameters2}
2\mu_3 = \lambda_3-\lambda_4\,,\;\;\;\;
2\mu_4 = \lambda_3+\lambda_4\,,
\end{equation}
then \cite{Marshakov:2009gn}:
\begin{eqnarray}
\label{limit1}
 \mathcal{B}_{\Delta}\! \left[^{ \Delta_3 \; \Delta_2}_{\Delta_4 \;\Delta_1} \right]\!(x)
&\begin{array}{c}
{\scriptstyle\mu_4 \to \infty}
    \\[-10pt]
    \longrightarrow
\\[-10pt]
 \scriptstyle \mu_4 x = \Lambda
\end{array}   &
\bra{\Delta,  \mu_3 ,{\textstyle \frac{1}{2}}\Lambda}V_{\Delta_2}(1)\ket{\Delta_1},
\\
\label{limit2}
 \mathcal{B}_{\Delta}\! \left[^{ \Delta_3 \; \Delta_2}_{\Delta_4 \;\Delta_1} \right]\!(x)
&\begin{array}{c}
{\scriptstyle \mu_3,\,\mu_4 \to \infty}
    \\[-10pt]
    \longrightarrow
\\[-10pt]
 \scriptstyle \mu_3\mu_4 x = \Lambda^2
\end{array}   &
\bra{\Delta,\Lambda^2}V_{\Delta_2}(1)\ket{\Delta_1},
\\
\label{limit3}
 \mathcal{B}_{\Delta}\! \left[^{ \Delta_3 \; \Delta_2}_{\Delta_4 \;\Delta_1} \right]\!(x)
&\begin{array}{c}
{\scriptstyle \mu_1,\,\mu_4 \to \infty}
    \\[-10pt]
    \longrightarrow
\\[-10pt]
 \scriptstyle \mu_1\mu_4 x = \Lambda^2
\end{array}   &
\left\langle \Delta, \mu_2,{\textstyle \frac{1}{2}}\Lambda\, |\, \Delta, \mu _3, {\textstyle \frac{1}{2}}\Lambda\right\rangle,
\\
\label{limit4}
 \mathcal{B}_{\Delta}\! \left[^{ \Delta_3 \; \Delta_2}_{\Delta_4 \;\Delta_1} \right]\!(x)
&\begin{array}{c}
{\scriptstyle \mu_1,\,\mu_2,\,\mu_4 \to \infty}
    \\[-10pt]
    \longrightarrow
\\[-10pt]
 \scriptstyle \mu_1\mu_2\mu_4 x = \Lambda^3
\end{array}   &
\left\langle \Delta,  \mu_3 ,{\textstyle \frac{1}{2}}\Lambda\, |\,\Delta, \Lambda^2 \right\rangle,
\\
\label{limit5}
\mathcal{B}_{\Delta}\! \left[^{ \Delta_3 \; \Delta_2}_{\Delta_4 \;\Delta_1} \right]\!(x)
&\begin{array}{c}
{\scriptstyle \mu_1,\,\mu_2,\,\mu_3,\,\mu_4 \to \infty}
    \\[-10pt]
    \longrightarrow
\\[-10pt]
 \scriptstyle \mu_1\mu_2\mu_3\mu_4 x = \Lambda^4
\end{array}   &
\left\langle \Delta,  \Lambda^2\, |\,\Delta, \Lambda^2 \right\rangle.
\end{eqnarray}
As it was demonstrated in \cite{Poghossian:2009mk}
the recursive relations for the irregular blocks  can be derived by analyzing
decoupling limits of Zamolodchikov's recursive formula for the elliptic 4-point block
\cite{Zamolodchikov:ie,Zamolodchikov:2,Zamolodchikov:3}
\begin{equation}
\label{HH}
{\cal H}_{\Delta}\!\left[_{\Delta_{4}\;\Delta_{1}}^{\Delta_{3}\;\Delta_{2}}\right]\!(\,q)
=  1 + \sum_{n=1}^\infty
(16q)^{\;n} H^{\,n}_{\Delta}\!\left[_{\Delta_{4}\;\Delta_{1}}^{\Delta_{3}\;\Delta_{2}}\right]\
\end{equation}
defined by
\begin{equation}
\label{prefac}
\mathcal{B}_{\Delta}\! \left[^{ \Delta_3 \; \Delta_2}_{\Delta_4 \;\Delta_1} \right]\!(x)
=
\left( \frac{x}{16q}\right)^{\lambda^2 \over 4} (1-x)^{\frac{Q^2}{4}-\Delta_1-\Delta_3}
[\theta_3(q)]^{3Q^2 -4(\Delta_1+\Delta_2+\Delta_3+\Delta_4)} \,
\mathcal{H}_{\Delta}\! \left[^{ \Delta_3 \; \Delta_2}_{\Delta_4 \;\Delta_1} \right]\!(q)
\end{equation}
where
$$
\theta_3(q)=\sum\limits_{-\infty}^{\infty}q^{n^2},
\;\;\;\;\;\;
q(x) = {\rm e}^{-\pi {K(1-x)\over K(x)}},
\;\;\;\;\;\;
K(x)=\int\limits_{0}^1{dt\over \sqrt{(1-t^2)(1-xt^2)}}\,.
$$
The coefficients in (\ref{HH})
are uniquely determined by Zomolodchikov's recursive formula:
\begin{eqnarray}
\label{recrelHH}
H^{\,n}_{\Delta}\!\left[_{\Delta_{4}\;\Delta_{1}}^{\Delta_{3}\;\Delta_{2}}\right]
&=&\delta_{n,0}+
\sum_{
1 \leqslant rs \leqslant n}
\frac{
 A_{rs}\,\prod\limits_{i=1}^{4}Y_{rs}(\mu_{i})}
{\Delta-\Delta_{rs}}
\;H^{\,n-rs}_{\Delta_{rs} +rs}\!
\left[_{\Delta_{4}\;\Delta_{1}}^{\Delta_{3}\;\Delta_{2}}\right],
\end{eqnarray}
where $\mu_i$ are defined by (\ref{parameters1}),(\ref{parameters2}) and
\begin{eqnarray}
\label{delta:rs}
\Delta_{rs}
&=&\frac{Q^2}{4}- \frac14 \left(r b + {s} b^{-1}\right )^2,
\\
\label{A}
A_{rs}
&=&
{\textstyle\frac12}
\hspace{-10pt}
\prod_{\begin{array}{c}\\[-24.5pt]\scriptstyle p=1-r\\[-6pt]\hspace{-8pt}\scriptstyle (p,q)\neq
\end{array}}^{r}
\hspace{-15pt}\prod_{\begin{array}{c}\\[-24.5pt]\scriptstyle q=1-s\\[-6pt]\hspace{0pt}\scriptstyle  (0,0),(r,s)
\end{array}}^{s}
\hspace{-10pt}{1\over pb +q b^{-1}}\,,
\\
\label{Y:factor}
Y_{rs}(\mu)
&=&
\hskip -5pt
\prod\limits_{\textstyle {}^{\hskip 10pt p=1-r}_{p+r=1\,{\rm mod}\,2}}^{r-1}\
\prod\limits_{\textstyle {}^{\hskip 10pt q=1-s}_{q+s=1\,{\rm mod}\,2}}^{s-1}
\left(\mu - \frac{pb + qb^{-1}}{2}\right).
\end{eqnarray}
In the limits $\mu_4\to \infty,\ \mu_4 x=\Lambda$ and $\mu_3\mu_4\to \infty,\ \mu_3\mu_4 x=\Lambda^2$
one has
$$
16 \mu_4 q(x) \longrightarrow \mu_4 x = \Lambda\,,\;\;\;\;\;\;
16\mu_3\mu_4 q(x) \longrightarrow \mu_3\mu_4 x = \Lambda^2\,,
$$
and the limits of (\ref{prefac}) take the form
\begin{eqnarray*}
\bra{\Delta,  \mu_3 ,{\textstyle \frac{1}{2}}\Lambda}V_{\Delta_2}(1)\ket{\Delta_1}
&=&
\exp\left( -{\textstyle \frac{1}{64}}\Lambda^2 - {\textstyle\frac12} \mu_3 \Lambda \right)
\left(1+\sum\limits_{n=1}^\infty H_n(\Delta,\mu_1,\mu_2,\mu_3)\Lambda^n\right),
\\
\bra{\Delta,\Lambda^2}V_{\Delta_2}(1)\ket{\Delta_1}
&=&
\exp\left( -{\textstyle \frac{1}{2}}\Lambda^2\right)
\left(1+\sum\limits_{n=1}^\infty H_n(\Delta,\mu_1,\mu_2)\Lambda^{2n}\right).
\end{eqnarray*}
Since
$
\lim\limits_{\mu \to \infty} \mu^{-rs}Y_{rs}(\mu)=1
$
the coefficients $H_n(\Delta,\mu_1,\mu_2,\mu_3)$ and
$H_n(\Delta,\mu_1,\mu_2)$ satisfy the recursive relations:
\begin{eqnarray}
\nonumber
\label{recursion3}
 H_n(\Delta,\mu_1,\mu_2,\mu_3)
 & = &
\delta^n_0 + \hskip -5pt
\sum\limits_{
1 \leqslant rs \leqslant n}
\hskip -2pt \frac{
 A_{rs}Y_{rs}(\mu_{1})Y_{rs}(\mu_{2})Y_{rs}(\mu_{3})}
{\Delta-\Delta_{rs}}
H_{n-rs}(\Delta_{rs}+rs,\mu_1,\mu_2,\mu_3)\,,
\\
\label{recursion2b}
 H_n(\Delta,\mu_1,\mu_2)
 & = &
\delta^n_0 + \hskip -5pt
\sum\limits_{
1 \leqslant rs \leqslant n}
\hskip -2pt \frac{
 A_{rs}Y_{rs}(\mu_{1})Y_{rs}(\mu_{2})}
{\Delta-\Delta_{rs}}\,
H_{n-rs}(\Delta_{rs}+rs,\mu_1,\mu_2)\,.
\end{eqnarray}
In the other cases (\ref{limit3}), (\ref{limit4}), (\ref{limit5}) the limit of the prefactor in
(\ref{prefac}) is simply 1 and one gets
\begin{eqnarray}
\label{recursion2a}
 \left\langle \Delta, \mu_1, n \, |\,\Delta, \mu_2,n \right\rangle
 & = &
\delta^n_0
\\
\nonumber
&&\hspace{-60pt}+ \hskip -5pt
\sum\limits_{
1 \leqslant rs \leqslant n}
\hskip -2pt \frac{   A_{rs} Y_{rs}(\mu_1) Y_{rs}(\mu_2)} {
\Delta-\Delta_{rs} }\,
 \left\langle \Delta_{rs}+rs, \mu_1, n-rs \, |\,\Delta_{rs}+rs,\mu_2, n-rs \right\rangle,
\\
\label{recursion1}
 \left\langle \Delta, \mu, n \, |\,\Delta, n \right\rangle
 & = &
\delta^n_0
\\
\nonumber
&&\hspace{-60pt}+ \hskip -5pt
\sum\limits_{
1 \leqslant rs \leqslant n}
\hskip -2pt \frac{   A_{rs} Y_{rs}(\mu)} {
\Delta-\Delta_{rs} }\,
 \left\langle \Delta_{rs}+rs, \mu, n-rs \, |\,\Delta_{rs}+rs, n-rs \right\rangle,
\\
\label{recursion0}
 \left\langle \Delta,  n \, |\,\Delta, n \right\rangle
 & = &
\delta^n_0 + \hskip -5pt
\sum\limits_{
1 \leqslant rs \leqslant n}
\hskip -2pt \frac{ A_{rs}} {
\Delta-\Delta_{rs} }\,
 \left\langle \Delta_{rs}+rs,  n-rs \, |\,\Delta_{rs}+rs, n-rs \right\rangle.
\end{eqnarray}
Comparing (\ref{recursion2b}) and (\ref{recursion2a}) one obtains
the equivalence of two different realizations of the irregular block with two
$\mu$ parameters proposed in \cite{Gaiotto:2009ma}:
$$
\bra{\Delta,\Lambda^2}V_{\Delta_2}(1)\ket{\Delta_1}=
{\rm e}^{-{\Lambda^2\over 2}}\left\langle \Delta, \mu _1, {\textstyle\frac{1}{2}}\Lambda \, \Big|\,
\Delta, \mu _2, {\textstyle\frac{1}{2}}\Lambda\right\rangle.
$$
This completes the derivation of the recursive relations required for the proof of the AGT conjecture.

We close this section by some remarks on  $Y_{rs}$ polynomials. They show up in the derivation
of the Zamolodchikov recursive relation in the factorization formula for the fusion polynomial
\cite{Hadasz:2006sb}\footnote{In ref. \cite{Hadasz:2006sb} the case of $N=1$ SCFT
is considered but the reasoning is the CFT case is essentialy the same.}
\begin{equation}
\label{nullvector}
\bra{\Delta_{rs}}O_{rs}^\dagger V_{\Delta_2}(1)\ket{\Delta_1}=(-1)^{rs}Y_{rs}(\mu_1)Y_{rs}(\mu_2)
\end{equation}
where as before the relation between parameters is given by (\ref{parameters1}) and $O_{rs}$
denotes
the combination of the Virasoro algebra generators
creating  the singular state  of level $rs$ out of the degenerate vacuum $\ket{\Delta_{rs}}$ normalized
by the condition that the coefficient in front of $L_{-1}^{rs}\ket{\Delta_{rs}}$ is equal 1.
Let us note that factorization formula (\ref{nullvector})
is a direct consequence of the null vector decoupling theorem
\cite{Teschner:2001rv}.

Another  interpretation of $Y_{rs}$ can be obtained by the derivation of the recursive
formulae for irregular blocks directly from their expressions in terms of the inverse Gram matrix.
Let us consider the block
\begin{eqnarray*}
\left\langle \Delta,  \mu ,{\textstyle \frac{1}{2}}\Lambda\, |\,
\Delta, \Lambda^2 \right\rangle
&=&
\sum_n \Lambda^{3n} 2^{-n}
\ \sum_{p=0}^{n/2} (-1)^{p}  (2\mu )^{n-2p}  \left[B^{n}_{c, \Delta}\right]^{[1^{n-2p},2^p],[1^n]}.
\end{eqnarray*}
In the generic case the only singularities  of $\left[ B^{\,{n}}_{c,\Delta}\right]^{M,N}$
as a function of $\Delta$ are simple poles at zeros of the Kac determinant
$
\Delta_{rs}
= \frac{Q^2}{4}- \frac14 \left(r b + {s} b^{-1}\right )^2
$
, $r\geqslant  1,\;s\geqslant 1, n\geqslant rs\geqslant 1$.
Since the degree of each minor of the Gram matrix as a function of $\Delta$ is strictly lower than the
degree of the Kac determinant itself there are no regular terms in the expansion:
\begin{eqnarray}
 \label{expansion_delta}
 \left\langle \Delta,\mu , n | \Delta, n  \right\rangle
&=&
 \delta_{n,0}+\sum_{1 \leq rs \leqslant n} \frac{\mathcal{R}^\mu _{rs,n}}{\Delta - \Delta_{rs}}\
\end{eqnarray}
except $n=0$. For the residue calculation it is convenient to choose a specific basis
in each subspace ${\cal V}^n_{c,\Delta}\subset {\cal V}_{c,\Delta}$
of level $n\geqslant rs$  formed by  vectors:
$$
L_{-K}O_{rs}\ket{\Delta}\,,\;\;\;\;|K|= n-rs\,,
$$
 and by
an arbitrary basis in the orthogonal complement of ${\rm Span}\{L_{-K}O_{rs}\ket{\Delta}\}$.
Due to the singular behavior  of the Gram matrix in the limit $\Delta\to \Delta_{rs}$
\cite{Hadasz:2009db,Hadasz:2006sb}
one has at the residue
\begin{eqnarray}
\nonumber
&& \hskip -1cm  \lim_{\Delta\to \Delta_{rs}} (\Delta-\Delta_{rs})\,
 \left\langle \Delta, \mu ,  n \, \Big|\,\Delta, n \right\rangle\,=
\\
\nonumber
&=&
 \lim_{\Delta\to \Delta_{rs}}(\Delta-\Delta_{rs})
 \sum_{\scriptscriptstyle |K|=|M|=n-rs}
\bra{ \Delta,   \mu , n}  L_{-K} O_{rs}\ket{\Delta }
\left[G^{n-rs}_{c, \Delta}\right]^{K,M}
\bra{\Delta} O_{rs}^\dagger L_{M}\ket{\Delta, n }
\\
\nonumber
&=&\lim_{\Delta\to \Delta_{rs}}(\Delta-\Delta_{rs}) \sum_{\scriptscriptstyle |K|=n-rs} \sum_{\scriptscriptstyle |J|=n}
 \sum_{\scriptstyle p=0}^{n\over 2}
\\ \nonumber
&&\, (2\mu )^{n-2p}\, (-1)^{p} 2^{-n} \,
\left[G^{n}_{c, \Delta}\right]^{J,[1^{n-2p},2^p]}
\bra{ \Delta } L_{J}L_{-K} O_{rs}\ket{\Delta }
\left[G^{n-rs}_{c, \Delta}\right]^{K,[1^{n-rs}]}
\\
\label{res1}
&&=A_{rs} \,
 \sum_{\scriptstyle p=0}^{n\over 2} \sum_{\scriptstyle q=0}^{\frac{n-rs}{2}} \, (2\mu )^{n-2p}\, (-1)^{p} 2^{-n}\,
e^{[1^{n-2p},2^p]}_{[1^{n-rs-2q},2^q]}
\left[G^{n-rs}_{c, \Delta_{rs}+rs}\right]^{[1^{n-rs-2q},2^q],[1^{n-rs}]}
\end{eqnarray}
where
$$
A_{rs}
\; = \;
\lim_{\Delta\to\Delta_{rs}}
\left(\frac{\bra{\Delta}O_{rs}^\dagger O_{rs}\ket{\Delta }}{\Delta - \Delta_{rs}(c)}
\right)^{-1}
$$
and $  e^{I}_{K}$ are the coefficients of the state $L_{-K}O_{rs}\ket{\Delta_{rs}}$
in the standard basis of $\mathcal{V}_{\Delta_{rs}}^{|K|+rs}$:
$$
L_{-K}O_{rs} \ket{\Delta_{rs}}= \sum_{|I|=|K|+rs} e^{I}_{K} L_{-I}\ket{\Delta_{rs}}.
$$
For our  normalization of $O_{rs} \ket{\Delta_{rs}}$
the exact form (\ref{A}) of the coefficient $A_{rs}$  was first proposed by Al.~Zamolodchikov in \cite{Zamolodchikov:ie} and
then justified  in
\cite{Zamolodchikov:2003yb}.
In formula (\ref{res1}) only the coefficients corresponding to the states
 generated from $\ket{\Delta_{rs}}$  by the operators $L_{-2}, L_{-1}$ are present.
  Let us define such coefficients for the singular state:
$$
O_{rs}\ket{\Delta_{rs}}= \sum_{k=0}^{\frac{rs}{2}}\, c^{(rs)}_k \,  L_{-1}^{rs-2k}L_{-2}^k\ket{\Delta_{rs}} + \ldots
$$
Then for an arbitrary $q$, the sum over $p$ in (\ref{res1}) is given in terms of $c^{(rs)}_p$ and $\mu $:
\begin{eqnarray}
\nonumber
X_{rs}(\mu ) &=&2^{-rs}
 \sum_{\scriptstyle p=q}^{{rs\over 2}+q}
(2\mu )^{rs-2(p-q)}\, (-1)^{p-q} \,
e^{[1^{n-2p},2^p]}_{[1^{n-rs-2q},2^q]}
\\
\nonumber\label{Xpoly}
 &=&
2^{-rs}\sum_{p=0}^{\frac{rs}{2}} (2\mu )^{rs-2p} \, (-1)^{p} \, c^{(rs)}_p.
\end{eqnarray}
This yields  the factorization formula for the residues ${\cal R}_{rs,n}^\mu $ in (\ref{expansion_delta}):
\begin{eqnarray*}
\lim_{\Delta\to \Delta_{rs}} (\Delta-\Delta_{rs})\,
 \left\langle \Delta, \mu,  n \, \Big|\,\Delta, n \right\rangle\,&=&
  A_{rs} X_{rs}(\mu)
\left\langle \Delta_{rs}+rs, \mu,  n-rs \, \Big|\,\Delta_{rs}+rs, n-rs \right\rangle.
\end{eqnarray*}
Expansion  (\ref{expansion_delta}) and the  formula above yield the
recursive relation:
\begin{eqnarray}
\nonumber\label{recursion2}
 \left\langle \Delta, \mu, n \, \Big|\,\Delta, n \right\rangle
 & = &
\delta^n_0
\\
\nonumber
&&\hspace{-50pt}+ \hskip -5pt
\sum\limits_{
1 \leqslant rs \leqslant n}
\hskip -2pt \frac{   A_{rs} X_{rs}(\mu)} {
\Delta-\Delta_{rs} }\,
 \left\langle \Delta_{rs}+rs, \mu, n-rs \, \Big|\,\Delta_{rs}+rs, n-rs \right\rangle.
\end{eqnarray}
Comparing with (\ref{recursion1})
one gets
$
X_{rs}(\mu) = Y_{rs}(\mu)$ which provides the factorization formula for
the $X_{rs}$ polynomial:
\begin{equation}
\label{X-P}
2^{-rs}\sum_{p=0}^{\frac{rs}{2}} (2\mu )^{rs-2p} \, (-1)^{p} \, c^{(rs)}_p\; =\;
2^{-rs}\!\!\!\!\!\!\!\!\prod\limits_{\begin{array}{c} \\[-22pt]\scriptstyle k=1-r\\[-6pt]\scriptstyle k+r=1\,  {\rm mod}\, 2
\end{array}}^{r-1}
\!\!\!
\prod\limits_{\begin{array}{c}\\[-22pt]\scriptstyle l=1-s\\[-6pt]\scriptstyle l+s=1\, {\rm  mod}\, 2
\end{array}}^{s-1}
\!\!\!\!\!\!\!\!\!{( 2\mu - k b - l b^{-1} )} \
\end{equation}
and implies an unexpected (from the point of view of the original definition of $X_{rs}$) relation with
the fusion polynomial (\ref{nullvector}):
$$
\bra{\Delta_{rs}}O_{rs}^\dagger V_{\Delta_2}(1)\ket{\Delta_1}=(-1)^{rs}X_{rs}(\mu_1)X_{rs}(\mu_2)
\ .
$$

\section{Recursive relations for the Nekrasov partition functions}
We shall  discuss the instanton contribution to the Nekrasov partition function of the
${\cal N}=2$ supersymmetric, U(2) gauge theory with $N_{\!f}$ hypermultiplets in the antifundamental
representation \cite{Nekrasov:2002qd}. It can be written as a sum over pairs of Young diagrams,
\begin{eqnarray}
\label{Z:gauge}
Z^{N_{\!f}}({\sf p}_\alpha,\mu_{\!f},b,\hbar;q) &=& 1+\sum\limits_{N=1}^{\infty}
Z^{N_{\!f}}_N({\sf p}_1,{\sf p}_2,\mu_{\!f},b)
\left(q\hbar^{N_f-4}\right)^N,
\\
\label{charge:N:contribution}
Z^{N_{\!f}}_N({\sf p}_1,{\sf p}_2,\mu_{\!f},b)
&=&
 \sum\limits_{|\vec Y|=N}\hskip -3pt Z^{N_{\!f}}({\sf p}_\alpha,\mu_{\!f},b;\vec Y)\,,
\end{eqnarray}
where\footnote{Our notation is close to the one in \cite{Fateev:2009aw}.}
 $\hbar {\sf p}_\alpha = a_\alpha,\ \alpha=1,2$ are the vev-s of the scalar component of the ${\cal N} = 2$ gauge
supermultiplet,
 $\hbar b = \epsilon_1,\ \hbar b^{-1} = \epsilon_2$ are the parameters of the $\Omega$ background,
 $Q=b+b^{-1}$,
$\hbar\left(\mu_f+{Q\over 2}\right) = m_{\!f},\ f = 1,\ldots N_{\!f}$
are the mass parameters of the hypermultiplets
and $|\vec Y|$ denotes the total number of boxes in the pair of Young diagrams $\vec Y = (Y_1,Y_2).$

The contribution to the partition function parameterized by a specific pair of Young diagrams is of the form \cite{Bruzzo:2002xf}
\begin{equation}
\label{Nakrasov:gauge}
Z^{N_{\!f}}\!({\sf p}_\alpha,\mu_{\!f},b;\vec Y) =\!
\Bigg(
\prod\limits_{\alpha=1}^2\prod\limits_{\langle m,n\rangle \in Y_\alpha}\hskip -5pt S_{\alpha}(\langle m,n\rangle)\!
\Bigg)\!
\Bigg(
\prod\limits_{\alpha,\beta =1}^2\prod\limits_{\langle m,n\rangle \in Y_\alpha}
\frac{1}{E_{\alpha\beta}(\langle m,n\rangle)\big(Q-E_{\alpha\beta}(\langle m,n\rangle)\big)}
\Bigg)
\end{equation}
where
\begin{eqnarray*}
S_\alpha(\langle m,n\rangle)
& = &
\prod\limits_{f=1}^{N_{\!f}}\left({\sf p}_{\alpha} + (m-1)b + (n-1)b^{-1} + \mu_{\!f}+{\textstyle{Q\over 2}}\right),
\\
E_{\alpha\beta}(\langle m,n\rangle)
& = &
{\sf p}_\alpha - {\sf p}_\beta - bH_{Y_\beta}(\langle m,n\rangle) + b^{-1} \left(V_{Y_\alpha}(\langle m,n\rangle)+ 1\right).
\end{eqnarray*}
The $N-$box diagram $Y$ can be described by an ordered sequence of  natural numbers $k_1 \geq k_2 \ldots \geq k_l > k_{l+1} = 0$
corresponding to the heights of columns of $Y$.
The vertical distance from the edge of the diagram of the box $\langle m,n\rangle$
situated in the $n-$th row (counted from the lowest one) of the $m-$th column (counted from the left)  is equal to
\[
V_Y(\langle m,n\rangle) = k_m(Y) - n,
\]
while  the horizontal distance  reads
\[
H_Y(\langle m,n\rangle) = k_n(Y^{\rm\scriptscriptstyle T}) - m
\]
where $Y^{\rm\scriptscriptstyle T}$ denotes the transposed diagram.

The contribution to the partition functions from instantons of topological charge
$N$ can be expressed as a contour integral \cite{Nekrasov:2002qd}
\begin{eqnarray}
\label{Nekrasov:1}
Z^{N_{\!f}}_N({\sf p}_1,{\sf p}_2,\mu_{\!f},b)
& = &
\frac{Q^N}{N!}
\oint\limits_{\mathbb R}\!\frac{d\phi_N}{2\pi i}\ldots \oint\limits_{\mathbb R}\!\frac{d\phi_1}{2\pi i}
\Bigg(
\prod\limits_{k=1}^N\frac{{\cal Q}_{\!f}(\phi_k)}{{\cal P}(\phi_k-i0){\cal P}(\phi_k+Q+i0)}
\\
\nonumber
&&
\hspace*{3.4cm}
\times
\prod\limits_{{}^{i,j =1}_{\hskip 3pt i\neq j}}^N\frac{\phi_{ij}
\big(\phi_{ij}-Q\big)}{\big(\phi_{ij}-b - i0\big)\big(\phi_{ij}-b^{-1}-i0\big)}
\Bigg)
\end{eqnarray}
where $Q = b + b^{-1},$ $\phi_{ij} = \phi_i-\phi_j$ and
\[
{\cal P}(\phi) = (\phi-{\sf p}_1)(\phi-{\sf p}_2),
\hskip 1cm
{\cal Q}_{\!f}(\phi) = \prod_{f=1}^{N_{\!f}}(\phi+\mu_f+{\textstyle{Q\over 2}}).
\]
A contribution to the Nekrasov partition function parameterized by a pair of Young diagrams $(Y_1,Y_2)$
corresponds to a specific choice of the integration contours in  (\ref{Nekrasov:1})
\cite{Fateev:2009aw} (see also the Appendix).
If the box $\langle r,s\rangle$  belongs to the diagram $Y_\alpha,$ then for some $k$
the contour of integration over $\phi_k$ surrounds only the pole
at ${\sf p}_\alpha + (r-1)b + (s-1)b^{-1},$ yielding the corresponding residue.
This pole is present in the integrand if and only if
at least $rs-1$ integrals were already computed and the contributions from all the poles at ${\sf p}_\alpha + (m-1)b + (n-1)b^{-1}$
with $m \leq r,\ n\leq s, (m,n)\neq(r,s)$ (one pole per one integral) were taken  into account.
If we visualize the computation of the contribution corresponding to a pair $(Y_1,Y_2)$ as ``building''
the  Young diagrams by adding subsequent boxes one by one,
than one can  add a  box $\langle r,s\rangle$
only if all the boxes in the rectangular $1\leq m \leq r, \ 1 \leq n \leq s$
save the right upper corner $\langle m,n\rangle=\langle r,s\rangle$ are already present.

It was demonstrated in
\cite{Fateev:2009aw} that the poles of $Z^{N_{\!f}}_N({\sf p}_1,{\sf p}_2,\mu_{\!f},b)$ appear solely at ${\sf p}_{12} \equiv {\sf p}_1-{\sf p}_2=
\mp(rb +sb^{-1}),$
with $1\leq r,s \leq N,\ rs \leq N.$ Moreover, these and only these pairs of diagrams which include as a subset of $Y_1$ the rectangle
$1\leq m \leq r,\ 1 \leq n \leq s$ contribute to the pole at ${\sf p}_{12} = -rb-sb^{-1}$ and only those which include
the rectangle $1\leq m \leq r,\ 1 \leq n \leq s$ as a subset of $Y_2$ contribute to the pole at ${\sf p}_{21} = -rb-sb^{-1}.$

Let us calculate the residue of the pole at ${\sf p}_{12} = -rb - sb^{-1}.$
One has to take into account only those contributions in  (\ref{Nekrasov:1})  for which $rs$ out of $N$ integrals
 are evaluated by calculating the residues at ${\sf p}_1, {\sf p}_1 +b,\ldots, {\sf p}_1 + (r-1)b + (s-1)b^{-1}$.
All the other contributions are finite in the limit ${\sf p}_{12} = -rb - sb^{-1}$.
By a suitable re-labeling of indices (which yields a combinatorial factor $\left(^{ N}_{rs}\right)$)
we may denote by $\phi_{mn}$ the variable of integration in the integral evaluated by taking a residue
at ${\sf p}_1 + (m-1)b + (n-1)b^{-1}.$
If we declare the indices $i,j$ and $k,l$ to satisfy
\[
1 \leq i, j \leq rs,
\hskip 1cm
rs < k,l \leq N,
\]
then the residue of $Z_N({\sf p}_1,{\sf p}_2,\mu_{\!f},b)$ at ${\sf p}_{12} = -rb - sb^{-1}$ is equal to
\begin{eqnarray*}
{\rm Res}\, Z^{N_{\!f}}_N({\sf p}_1,{\sf p}_2,\mu_{\!f},b)
&=&
\frac{Q^{N-rs}}{(N-rs)!}
\int\limits_{\mathbb R}\!\frac{d\phi_N}{2\pi i}\ldots \int\limits_{\mathbb R}\!\frac{d\phi_{r s+1}}{2\pi i}
\prod\limits_{k\neq l}\frac{\phi_{kl} \big(\phi_{kl}-Q\big)}{\big(\phi_{kl}-b -i0\big)\big(\phi_{kl}-b^{-1}-i0\big)}
\\
&\times &\prod\limits_{k=rs+1}^N\frac{{\cal Q}_{\!f}(\phi_k)}{{\cal P}(\phi_k){\cal P}(\phi_k+Q)}
\frac{Q^{rs}}{(rs)!}
\,K_{rs}
\end{eqnarray*}
where
\begin{eqnarray*}
K_{rs}
&
=
&
 {\rm Res}
\int\limits_{\mathbb R}\!\frac{d\phi_{rs}}{2\pi i}\ldots \int\limits_{\mathbb R}\!\frac{d\phi_{1}}{2\pi i}\
\prod\limits_{k,i}\frac{\phi_{ki}^2 \big(\phi_{ki}^2-Q^2\big)}{\big(\phi_{ki}^2-(b-i0)^2\big)\big(\phi_{ki}^2-(b-i0)^{-2}\big)}
\\[6pt]
&\times&
\prod\limits_{i=1}^{rs}\frac{{\cal Q}_{\!f}(\phi_i)}{{\cal P}(\phi_i-i0){\cal P}(\phi_i+Q+i0)}\
\prod\limits_{i\neq j}\frac{\phi_{ij} \big(\phi_{ij}-Q\big)}{\big(\phi_{ij}-b- i0\big)\big(\phi_{ij}-b^{-1}-i0\big)}
\\[6pt]
&=&
{\rm Res}\,Z^{N_{\!f}}_N({\sf p}_1,{\sf p}_2,\mu_{\!f},b,\vec Y_{r,s})
\hskip -5pt
\prod\limits_{k=rs+1}^N
\prod\limits_{m=1}^{r}\prod\limits_{n=1}^{s}
\frac{(\phi_k-x_{mn})^2 \big((\phi_k-x_{ms})^2-Q^2\big)}{\big((\phi_k-x_{ms})^2-b^2\big)\big((\phi_k-x_{ms})^2-b^{-2}\big)}\,.
\end{eqnarray*}
In the last line, $Z^{N_{\!f}}_N({\sf p}_1,{\sf p}_2,\mu_{\!f},b,\vec Y_{rs})$ is a contribution to
the partition function corresponding to the pair $\vec Y_{rs}$
 such that $Y_1$ is a $r\times s$ rectangle and $Y_2 = \emptyset,$ while
$x_{mn} = {\sf p}_1 + (m-1)b + (n-1)b^{-1}.$

After some simple algebra we get
\begin{eqnarray}
\label{intermediate}
\nonumber
&&
\hskip -2cm
\frac{1}{{\cal P}(\phi_k){\cal P}(\phi_k+Q)}\
\prod\limits_{m=1}^{r}\prod\limits_{n=1}^{s}
\frac{(\phi_k-x_{mn})^2 \big((\phi_k-x_{mn})^2-Q^2\big)}{\big((\phi_k-x_{mn})^2-b^2\big)\big((\phi_k-x_{mn})^2-b^{-2}\big)}
\\[6pt]
& = &
\frac{1}{(\phi_{k}-{\sf p}_1-rb)(\phi_{k}+Q-{\sf p}_1-rb)(\phi_{k}-{\sf p}_1-sb^{-1})(\phi_{k}+Q-{\sf p}_1-sb^{-1})}
\\[6pt]
\nonumber
& \times &
\frac{(\phi_{k}-{\sf p}_1 -rb -sb^{-1})(\phi_{k}+Q-{\sf p}_1 -rb -sb^{-1})}{(\phi_{k}-{\sf p}_2)(\phi_{k}+Q-{\sf p}_2)}\,.
\end{eqnarray}
For ${\sf p}_{12} = -rb - sb^{-1}$ the factor in the last line of (\ref{intermediate}) is equal to 1 and
\begin{eqnarray}
\label{simplification}
\nonumber
&&
\hskip -2cm
\frac{1}{{\cal P}(\phi_k){\cal P}(\phi_k+Q)}\
\prod\limits_{m=1}^{r}\prod\limits_{n=1}^{s}
\frac{(\phi_k-x_{mn})^2 \big((\phi_k-x_{mn})^2-Q^2\big)}{\big((\phi_k-x_{mn})^2-b^2\big)\big((\phi_k-x_{mn})^2-b^{-2}\big)}\Big|_{{\sf p}_{12}= -rb-sb^{-1}}
\\[4pt]
&= &
\frac{1}{\widetilde{\cal P}(\phi_{k})\widetilde{\cal P}(\phi_{k}+Q)}
\end{eqnarray}
with
\begin{equation}
\label{P:tilde}
\widetilde{\cal P}(\phi) = (\phi-{\sf p}_1-rb)(\phi-{\sf p}_1-sb^{-1}).
\end{equation}
It leads to the  relation
\begin{eqnarray}
\label{recur:iterm}
{\rm Res}\, Z^{N_{\!f}}_N({\sf p}_1,{\sf p}_2,\mu_{\!f},b)
=
{\rm Res}\,Z^{N_{\!f}}({\sf p}_1,{\sf p}_2,\mu_{\!f},b;\vec Y_{rs})\:
Z^{N_{\!f}}_{N-rs}({\sf p}_1+rb, {\sf p}_1+sb^{-1},\mu_{\!f},b).
\end{eqnarray}
For the pair of diagrams $\vec Y_{rs}=(r\times s, \emptyset)$ one has
\begin{eqnarray*}
H_{Y_1}(\langle m,n\rangle) & = & r-m,
\hskip 1cm
H_{Y_2}(\langle m,n\rangle) \; = \;  -m,
\\
V_{Y_1}(\langle m,n\rangle) & = & s-n,
\hskip 1cm
V_{Y_2}(\langle m,n\rangle) \; = \;  -n,
\end{eqnarray*}
 and
\begin{eqnarray*}
E_{11}(\langle m,n\rangle)
& = &
(m-r)b + (s-n+1)b^{-1},
\\[6pt]
E_{12}(\langle m,n\rangle)
& = &
{\sf p}_{12} + mb+ (s-n+1)b^{-1}.
\end{eqnarray*}
The formula (\ref{Nakrasov:gauge}) thus  gives
\begin{eqnarray}
\label{A:factor:iterm}
&&
\hskip -3cm
{\rm Res}\!\!\prod\limits_{\alpha,\beta=1}^2\prod\limits_{\langle m,n\rangle \in Y_i}\!
\frac{1}{E_{\alpha\beta}(\langle m,n\rangle)(Q-E_{\alpha}(\langle m,n\rangle))}\Big|_{{\sf p}_{1\!2} = -rb -sb^{-1}}\; =
\\
\nonumber
&&
\hskip 2cm = \;
\prod\limits_{m=1-r}^r\prod\limits_{n=1-s}^s \frac{1}{mb + nb^{-1}},
\hskip 1cm \langle m,n\rangle \neq \langle 0,0\rangle.
\end{eqnarray}
Let us now assume
\[
{\sf p}_1 = -{\sf p}_2 = {\sf p}
\]
which corresponds to SU(2) rather than U(2) gauge group. The pole at ${\sf p}_{12} = -rb - sb^{-1}$ then corresponds to
\[
{\sf p} = -\frac12\left(rb + sb^{-1}\right).
\]
For the pair of diagrams $\vec Y_{rs} = (r\times s, \emptyset)$ we thus have
\begin{eqnarray*}
&&
\hskip -3cm
\prod\limits_{\alpha=1}^2\prod\limits_{\langle m,n\rangle \in Y_\alpha}
\hskip -5pt S_{\alpha}(\langle m,n\rangle)\Big|_{{\sf p} = -\frac{rb + sb^{-1}}{2}}
\\
& = &
\prod\limits_{f=1}^{N_{\!f}}\prod\limits_{m=1}^r\prod\limits_{n=1}^s
\left[\left(m-1-\textstyle{\frac12}r\right)b + \left(n-1 -\textstyle{\frac12}s\right)b^{-1} +  \mu_{\!f}+{\textstyle{Q\over 2}}\right]
\\
&=&
\prod\limits_{f=1}^{N_{\!f}}Y_{rs}(\mu_{\!f})
\end{eqnarray*}
where $Y_{rs}$ are defined by (\ref{Y:factor}).
Since  $Z_N({\sf p}_1,{\sf p}_2,\mu_{\!f},b)$ is  a symmetric function
of ${\sf p}_1-{\sf p}_2 = 2{\sf p}$
\[
Z_N({\sf p},\mu_{\!f},b) \equiv Z_N({\sf p},-{\sf p},\mu_{\!f},b) =   Z_N(-{\sf p},\mu_{\!f},b),
\]
the residues at the poles at ${\sf p}_{12} = -rb -sb^{-1}$ and ${\sf p}_{21} = -rb -sb^{-1}$ differ only
by a sign. Equations
(\ref{recur:iterm}) and (\ref{A:factor:iterm}) then yield for $4{\sf p}^2 \to \left(rb+sb^{-1}\right)^2$
\begin{equation}
\label{recur:iterm:2}
 Z_N^{N_{\!f}}({\sf p},\mu_{\!f},b) \; =
 \;- \frac{4 A_{rs}\,\prod\limits_{f=1}^{N_{\!f}}Y_{rs}(\mu_{\!f})}{4{\sf p}^2 -\left(rb+sb^{-1}\right)^2}\
 Z^{N_{\!f}}_{N-rs}({\textstyle {1\over 2}}(rb-sb^{-1}),\mu_{\!f},b) + {\cal O}(1)
\end{equation}
where
$
A_{rs}
$
is given by formula (\ref{A}).
For ${\sf p}\to \infty$  relation (\ref{Nakrasov:gauge}) implies
\[
 Z_N^{N_{\!f}}({\sf p},\mu_{\!f},b) = {\cal O}\left({\sf p}^{2N(N_{\rm f}-2)}\right),
\]
hence
\[
\lim_{p\to\infty} Z^{N_{\!f}}_N({\sf p},\mu_{\!f},b;q) = \delta_{0,N}\;\;{\rm for} \;\;N_f=0,1.
\]
Together with (\ref{recur:iterm:2}) this gives the recursion relations:
\begin{eqnarray*}
 Z_N^{0}({\sf p},b)
&=&
\delta_{0,N} - \sum\limits_{
1 \leqslant rs \leqslant N}
\frac{4A_{rs}}{4{\sf p}^2 -\left(rb+sb^{-1}\right)^2}\
 Z_{N-rs}^{0}\left({\textstyle {1\over 2}}(rb-sb^{-1}),b\right),
 \\
 Z_N^{1}({\sf p},\mu,b)
&=&
\delta_{0,N} - \sum\limits_{
1 \leqslant rs \leqslant N}
\frac{4A_{rs}\,Y_{rs}(\mu)}{4{\sf p}^2 -\left(rb+sb^{-1}\right)^2}\
 Z_{N-rs}^{1}\left({\textstyle {1\over 2}}(rb-sb^{-1}),b\right).
\end{eqnarray*}
Regarding the partition function as a function of the conformal dimension
$
\Delta = \frac14Q^2 - {\sf p}^2
$
rather than the function of ${\sf p},$ one gets the recursion formulae of the form:
\begin{eqnarray}
\label{recur:0}
 Z_N^{0}(\Delta,b)
&=&
\delta_{0,N} +  \sum\limits_{
1 \leqslant rs \leqslant N}
\frac{A_{rs}}{\Delta - \Delta_{rs}}
 Z_{N-rs}^{0}\left(\Delta_{rs}+rs,b\right),
 \\
\label{recur:1}
 Z_N^{1}(\Delta,\mu,b)
&=&
\delta_{0,N} + \sum\limits_{
1 \leqslant rs \leqslant N}
\frac{A_{rs}\,Y_{rs}(\mu)}{\Delta - \Delta_{rs}}
 Z_{N-rs}^{1}\left(\Delta_{rs}+rs,\mu,b\right),
\end{eqnarray}
where
$
\Delta_{rs} $ is given by relation (\ref{delta:rs}). Comparing with (\ref{recursion0}) and (\ref{recursion1})
one gets the AGT relation for $N_{\!f}=0,1$:
\begin{eqnarray*}
Z^{0}(\Delta,b,\hbar\,; \,\hbar^4\Lambda^4)&=&\left\langle \Delta, \Lambda^2  | \Delta,\Lambda^2  \right\rangle,
\\
Z^{1}(\Delta,\mu,b,\hbar\,; \,\hbar^3\Lambda^3)&=&\left\langle \Delta,  \mu ,{\textstyle \frac{1}{2}}\Lambda\, |\,
\Delta, \Lambda^2 \right\rangle.
\end{eqnarray*}

Calculating the large ${\sf p}$ asymptotic in the case $N_{\!f}=2$ is more involved.
Let us first note  that for  $\alpha\neq \beta$  and ${\sf p}={\sf p}_\alpha = -{\sf p}_\beta\to \infty:$
\begin{eqnarray*}
\frac{S_\alpha(\langle m,n\rangle)}{E_{\alpha\beta}(\langle m,n\rangle)\big(Q-E_{\alpha\beta}(\langle m,n\rangle)\big)}
& = & \frac{{\sf p}_\alpha^2}{-({\sf p}_\alpha-{\sf p}_\beta)^2} + {\cal O}({\sf p}^{-1})
\hskip 5mm \to \; - \frac14\,,
\end{eqnarray*}
hence, for $|\vec Y|=N$
\[
\prod\limits_{\textstyle{}^{\alpha,\beta = 1}_{\hskip 3pt\alpha\neq \beta}}^2\prod\limits_{\langle m,n\rangle \in Y_1}
\frac{S_\alpha(\langle m,n\rangle)}{E_{\alpha\beta}(\langle m,n\rangle)\big(Q-E_{\alpha\beta}(\langle m,n\rangle)\big)}\
\hskip 5mm \to \; \left(- \frac14\right)^N
\]
and
\begin{eqnarray*}
\lim_{{\sf p}\to\infty}Z^{2}({\sf p},\mu_1,\mu_2,b,\hbar;q)
\hskip -5pt
& = &
\hskip -5pt
\sum\limits_{N=0}^{\infty}\!
\left(-\frac{q}{4 \hbar^2}\right)^N\!
\sum\limits_{|\vec Y|=N}
\prod\limits_{\alpha=1}^2
\prod\limits_{\langle m,n\rangle \in Y_\alpha}
\frac{1}{E_{\alpha\alpha}(\langle m,n\rangle)\big(Q-E_{\alpha\alpha}(\langle m,n\rangle)\big)}
\\[6pt]
& = &
\hskip -3pt
\left(
\sum\limits_{N=0}^{\infty}\
\left(-\frac{q}{4 \hbar^2}\right)^N
\sum\limits_{|Y|= N}
\prod\limits_{\langle m,n\rangle \in Y}
\frac{1}{E_Y(\langle m,n\rangle)\big(Q-E_Y(\langle m,n\rangle)\big)}
\right)^2
\hskip -3pt,
\end{eqnarray*}
where
\[
E_Y(\langle m,n\rangle) = -b H_Y(\langle m,n\rangle) + b^{-1}(1+ V_Y(\langle m,n\rangle))\,.
\]
In order to calculate the sum
\begin{equation}
\label{asymptotics:2}
{\cal Z}_{N}(b)
=
\sum\limits_{|Y|= N}
\prod\limits_{\langle m,n\rangle \in Y}
\frac{1}{E_Y(\langle m,n\rangle)\big(Q-E_Y(\langle m,n\rangle)\big)}
\end{equation}
we shall use the  integral representation
\begin{equation}
\label{intrep:my}
{\cal Z}_{N}(b)
=
\frac{Q^N}{N!}
\oint\limits_{\mathbb R}\!\frac{d\phi_N}{2\pi i}\ldots \oint\limits_{\mathbb R}\!\frac{d\phi_1}{2\pi i}
\prod\limits_{k=1}^N\frac{1}{(\phi_k-i0)(\phi_k+Q+i0)}
\prod\limits_{{}^{i,j =1}_{\hskip 3pt i\neq j}}^N\frac{\phi_{ij} \big(\phi_{ij}-Q\big)}{\big(\phi_{ij}-b - i0\big)\big(\phi_{ij}-b^{-1}-i0\big)}.
\end{equation}
The relation (\ref{intrep:my}) can be obtained using  the results of  \cite{Moore:1998et} and \cite{Nakajima}
(formulae 6.4-6.12 in \cite{Moore:1998et}).
Applying  the reasoning presented in Appendix to the integral in (\ref{intrep:my}) one can show
 that the only singularities of ${\cal Z}_{N}(b)$ result from the collision of the poles at $\phi_k = (r-1)b + (s-1)b^{-1}$
(above the real axis) with the pole at $\phi_k = -Q$ (below the real axis), i.e.\ for
\[
rb + sb^{-1} = 0.
\]
Since $r$ and $s$ are positive integers  the only singularities of ${\cal Z}_{N}(b)$ can occur for
purely imaginary $b.$ However, for $b = i\beta,\ \beta \in {\mathbb R},$
\begin{eqnarray*}
&&
\hskip -2cm
E_Y(\langle m,n\rangle)
(Q-E_Y(\langle m,n\rangle)
\\[4pt]
& = &
\Big( 1+ V_Y(\langle m,n\rangle)+ \beta^2 H_Y(\langle m,n\rangle)\Big)
\Big(1+H_Y(\langle m,n\rangle) + \beta^{-2}V_Y(\langle m,n\rangle)\Big)
\end{eqnarray*}
is strictly positive. Thus the residue at the would-be pole  at $rb + sb^{-1}= 0$ actually vanishes and
${\cal Z}_{N}(b)$ is a holomorphic function of  $b.$
Now, for $b\to \infty$
\[
\frac{1}{E_Y(\langle m,n\rangle)\big(Q-E_Y(\langle m,n\rangle)\big)} \to 0
\]
unless $H_Y(\langle m,n\rangle) = 0,$ i.e.\ $Y$  consists solely of the first column and therefore
\[
\lim_{b\to\infty}
\sum\limits_{|Y|= N}
\prod\limits_{\langle m,n\rangle \in Y}
\frac{1}{E_Y(\langle m,n\rangle)\big(Q-E_Y(\langle m,n\rangle)\big)}
=
\prod\limits_{m=1}^{N} \frac{1}{1+ V_Y(\langle m,1\rangle)} = \frac{1}{N!}\,.
\]
The Liouville's boundedness theorem then implies that  ${\cal Z}_{N}(b)$ is a $b$ independent constant\footnote{
An independent check of this property can be made by  evaluating with a help of \cite{math}, Theorem 2.7, the value of ${\cal Z}_{N}(i)$}:
\[
\sum\limits_{|Y|= N}
\prod\limits_{\langle m,n\rangle \in Y}
\frac{1}{E_Y(\langle m,n\rangle)\big(Q-E_Y(\langle m,n\rangle)\big)}
=
\frac{1}{N!}\,.
\]
One thus gets
\begin{eqnarray*}
\lim_{{\sf p}\to\infty}Z^{2}({\sf p},\mu_1,\mu_2,b,\hbar;q)
& = &
{\rm e}^{-\frac12 {q\over \hbar^2 }}
\end{eqnarray*}
which along with (\ref{recur:iterm:2}) implies the  recursive relation
\begin{eqnarray}
\label{recur:2}
 Z_N^{2}(\Delta,\mu_1,\mu_2,b)
&=&
{(-1)^N\over 2^N N!}
\\
\nonumber
&+& \sum\limits_{
1 \leqslant rs \leqslant N}
\frac{A_{rs}\,Y_{rs}(\mu_1)Y_{rs}(\mu_2)}{\Delta - \Delta_{rs}}
 Z_{N-rs}^{2}\left(\Delta_{rs}+rs ,\mu_1,\mu_2,b\right).
\end{eqnarray}
Comparing with  recursive relations (\ref{recursion2b}) and (\ref{recursion2a}) one gets
the AGT correspondence for $N_{\!f}=2$
$$
Z^2(\Delta,\mu_1,\mu_2,b,\hbar;\, \hbar^2\Lambda^2)= \bra{\Delta,\Lambda^2}V_{\Delta_2}(1)\ket{\Delta_1}=
{\rm e}^{-{\Lambda^2\over 2}}\left\langle \Delta, \mu _1, {\textstyle\frac{1}{2}}\Lambda \, \Big|\,
\Delta, \mu _2, {\textstyle\frac{1}{2}}\Lambda\right\rangle.
$$

\section*{Acknowledgements}

The work of L.H. was  supported by MNII grant 189/6.PRUE/2007/7.

\appendix
\section*{Appendix. Singularities of the partition function}
\renewcommand{\theequation}{A.\arabic{equation}}
\setcounter{equation}{0}
In this appendix we shall  repeat  (a slightly elaborated version of) a proof of the following lemma due
to Fateev and Litvinov \cite{Fateev:2009aw}:

\noindent
{\bf Lemma:}
Let $f(\phi_1,\ldots,\phi_n)$ be a holomorphic function of all its arguments which grows for $\phi_i\to \infty$ slow enough to ensure
the convergence of the integral
\begin{equation}
\label{capF:definition}
F^{f}_N\!\left(\{a_\alpha\};\{\epsilon_a\}\right)
=
\int\limits_{\mathbb R}\!\frac{d\phi_1}{2\pi i}\cdots\!\int\limits_{\mathbb R}\!\frac{d\phi_N}{2\pi i}\
f(\phi_1,\ldots,\phi_N)
\prod\limits_{\alpha}\prod\limits_{i=1}^{N} \frac{1}{\phi_i-a_\alpha-0^{+}}\
\prod\limits_{a}\prod\limits_{{}^{i,j=1}_{\hskip 3pt i \neq j}}^{N}
\frac{1}{\phi_{ij}- \epsilon_a-0^{+}}\,.
\end{equation}
Then $F^{f}_N\left(\{a_\alpha\};\{\epsilon_a\}\right)$  is a holomorphic function of all $a_\alpha.$

\noindent
{\bf Proof.} Let us perform the integral over $\phi_N$ by closing the contour in the upper half-plane. With this choice
the poles at $\phi_N = a_\alpha$ and $\phi_N = \phi_j + \epsilon_a$ are inside the contour, while the poles at $\phi_N = \phi_j - \epsilon_a$ stay outside.
When we vary $a_\alpha$, some of these poles move and, when the positions of two poles coincide, a higher order singularity
of the integrand may occur.

If this happens for the poles on the same side of the integration contour, then we can move the contour away from such a colliding pair
and the integral stays finite. On the other hand, if the colliding poles ``squeeze'' the integration contour in between then we can deform
the contour of integration away only at a price of evaluating a residue (which can become singular in the limit) at one of the poles.

More precisely: take a pole at $\phi_N =\phi_j - \epsilon_b$ and deform the integration contour to the sum
of ${\cal C}$ and ${-\cal C}_{k,b},$ where
$\cal C$ encloses all the poles of $\phi_N$ in the upper half-plane {\bf and} the pole at $\phi_N =\phi_j - \epsilon_b,$ while ${\cal C}_{j,b}$ surrounds
just $\phi_N =\phi_j - \epsilon_b.$ By the argument above a singularity of the
integral over $\phi_N$  may appear
only from the integral over ${\cal C}_{j,b}.$ Adding such contributions for all $j< N$ and all  $b$ (and neglecting for the moment
all the factors which do not depend on $\phi_N$) we get:
{\small
\begin{eqnarray}
\label{the:new:poles}
\nonumber
&&
\hskip -8mm
{\cal I}_N = \sum\limits_{j,b}\oint\limits_{-{\cal C}_{k,b}}\!\frac{d\phi_N}{2\pi i}\
f(\phi_1,\ldots,\phi_N)
\prod\limits_{\alpha} \frac{1}{\phi_N-a_\alpha-0^{+}}\
\prod\limits_{a}\prod\limits_{i=1}^{N-1}
\frac{1}{\phi_{N}-\phi_i- \epsilon_a-0^{+}}\frac{1}{\phi_{N}-\phi_i- \epsilon_a+0^{+}}
\\[-6pt]
\\[-6pt]
\nonumber
&&
\hskip -2mm
=
\sum\limits_{j,b}f(\phi_1,\ldots,\phi_j-\epsilon_b)
\prod\limits_{\alpha} \frac{1}{\phi_j-a_\alpha- \epsilon_b-0^{+}}\ \frac{1}{2\epsilon_b}\!\!
\prod\limits_{{}^{\hskip 6pt a,\, i<N}_{(a,i)\neq (b,j)}}
\frac{1}{\phi_{ji} - \epsilon_a-\epsilon_b-0^{+}}\frac{1}{\phi_{ji} + \epsilon_a-\epsilon_b-0^{+}}.
\end{eqnarray}
}

There are indeed several new poles
on the right hand side of this equation. The first type is at $\phi_j = a_\alpha + \epsilon_b$
above the real axis. In order to expose the others let us note that  (\ref{the:new:poles}) is symmetric under
$(a,i)\leftrightarrow (b,j)$ so that  symmetrizing its right hand side
we get
{\small
\begin{eqnarray*}
{}\,2{\cal I}_N=\!
\sum\limits_{j,b}f(\phi_1,\ldots,\phi_j-\epsilon_b)
\prod\limits_{\alpha} \frac{1}{\phi_j-a_\alpha- \epsilon_b-0^{+}}\frac{1}{2\epsilon_b}\!\!\!
\prod\limits_{{}^{\hskip 6pt a, i<N}_{(a,i)\neq (b,j)}}\!\!\!\!
\frac{1}{\phi_{ij} + \epsilon_a+\epsilon_b+0^{+}}\frac{1}{\phi_{ij} - \epsilon_a+\epsilon_b+0^{+}}
\\[2pt]
\;+
\sum\limits_{i,a}f(\phi_1,\ldots,\phi_i-\epsilon_a)
\prod\limits_{\alpha} \frac{1}{\phi_i-a_\alpha- \epsilon_a-0^{+}} \frac{1}{2\epsilon_a}\!\!\!
\prod\limits_{{}^{\hskip 6pt b, j<N}_{(b,j)\neq (a,i)}}\!\!\!
\frac{1}{\phi_{ij} - \epsilon_a-\epsilon_b-0^{+}}\frac{1}{\phi_{ij} -\epsilon_a+ \epsilon_b-0^{+}}.
\end{eqnarray*}
}

It follows from the expression above that ${\cal I}_N$ has
poles above the real axis at
$\phi_i = \phi_j +\epsilon_a + \epsilon_b$ and poles below the real axis at $\phi_i = \phi_j -\epsilon_a - \epsilon_b.$
It seems that there is also a pole at $\phi_i = \phi_j +\epsilon_a -\epsilon_b.$
However, the residue of such a ``would be'' pole
contains a factor
\[
\frac{1}{2\epsilon_b}
\frac{1}{\phi_{ij} + \epsilon_a+\epsilon_b}
+
\frac{1}{2\epsilon_a}
\frac{1}{\phi_{ij} - \epsilon_a-\epsilon_b}
=
\frac{\epsilon_a+\epsilon_b}{2\epsilon_a\epsilon_b}
\frac{\phi_{ij}-\epsilon_a +\epsilon_b}{\phi_{ij}^2-(\epsilon_a+\epsilon_b)^2}
\]
and ${\cal I}_N$ is actually finite for $\phi_i \to \phi_j +\epsilon_a - \epsilon_b.$

After performing the integral over $\phi_N$ we arrive at the expression of the same structure as (\ref{capF:definition})
with $N-1$ instead of $N$ integrals, the function $f$ replaced with some other holomorphic function of $\phi_1,\ldots\phi_{N-1},$
and the sets $\{a_\alpha\}, \{\epsilon_a\}$ enlarged by adding points $\{a_\alpha+\epsilon_b\}, \{\epsilon_a+\epsilon_b\}$ with all
possible $\epsilon_b.$ After performing all but the last integral we arrive at the formula
\[
F^{f}_N\left(\{a_\alpha\};\{\epsilon_a\}\right)
=
\int\limits_{\mathbb R}\!\frac{d\phi_1}{2\pi i}\
\tilde f(\phi_1)\prod\limits_{\alpha}\prod\limits_{A}\frac{1}{\phi_1-a_\alpha - e_A- i0^+}
=
\prod\limits_{\alpha}\prod\limits_{A}\ \tilde f(a_\alpha + e_A),
\]
with $\{e_A\} = \{0,\epsilon_a,\epsilon_a+\epsilon_b,\ldots\}$ and some holomorphic function $\tilde f.$ The thesis of the lemma follows.

Let us now suppose that the function $f$ appearing in (\ref{capF:definition})
contains a
factor
\[
\prod\limits_{\nu}\prod\limits_{i=1}^N\frac{1}{\phi_i - \xi_\nu +i0^+}\,,
\]
with simple poles below the real axis. Adapting the reasoning given in the proof above we see that $F^{f}_N$
is no longer holomorphic. Its singularities  may only come from the collision of poles at $\phi_i = a_\alpha +e_A$ and
$\phi_i= \xi_\nu$ with the $\phi_i$ integration contour squeezed in between and are thus located at
\[
a_\alpha = \xi_{\nu} + e_A.
\]
In the situation of interest in  Section 3 we have
$\alpha=1,2,\ $ $\xi_1 = a_1-\epsilon$ and $\xi_2 = a_2-\epsilon.$ The poles in the variables $\phi_i$ can appear only at
\[
\phi_i = a_\alpha + (r-1)\epsilon_1 + (s-1)\epsilon_2, \hskip 1cm r,s \in {\mathbb Z}_{+}
\]
and the only singularities of the partition function as a function of $a_\alpha$ are the (simple in the
generic case) poles at
\[
a_1-a_2 = \mp(m\epsilon_1 + n\epsilon_2).
\]

\end{document}